\documentstyle[aps,preprint,floats,psfig]{revtex}
\tighten
\preprint{\scriptsize Invited talk given at MENU97, 7/28--8/1/97, Vancouver, B.C., 
Canada}

\begin{document}
\draft

\title{Gauge-invariant theory of pion photoproduction\\
 with dressed hadrons}
\author{Helmut Haberzettl\thanks{E-mail: {\tt helmut@gwis2.circ.gwu.edu}}\\
{\em Center for Nuclear Studies, Department of Physics\\ 
The George Washington University, Washington, D.C. 20052, U.S.A.}}
\maketitle

\begin{abstract}
Based on an effective field theory of hadrons in which quantum
chromodynamics (QCD) 
is assumed to provide the necessary bare cutoff functions, a
gauge-invariant theory of pion photoproduction with fully dressed nucleons
is described. The formalism provides consistent dynamical descriptions of 
$\pi N\rightarrow \pi N$ scattering and $\gamma N\rightarrow \pi N$
production mechanisms in terms of nonlinear integral equations for fully
dressed hadrons. The dressed hadron currents and the pion
photoproduction current satisfy gauge invariance in
a self-consistent manner. Approximations are discussed that make the
nonlinear formalism manageable in practice and yet preserve gauge invariance.
\pacs{}
\end{abstract}

\section*{INTRODUCTION}
It is the purpose of the present contribution to describe a comprehensive
theory for the production of pions due to the interactions
of incident photons with nucleons \cite{hh97}. The history of 
such descriptions goes
back to the fifties, and indeed many of the more general basic relations
have been well-known for about forty years (see, e.g.,  
\cite{ward50,kroll54,takah57,kazes59} and references therein). In recent years, the attention has
focused on approaches attempting to take into account the fact that all
hadrons involved in the reaction have an internal structure 
\cite{gross,afnan1,ohta,friar,coester,afnan,banerjee}.
The work reported here adds to the latter approaches by
providing a detailed theoretical framework for the gauge-invariant interactions
of physical --- i.e., fully dressed --- hadrons with photons. 
The description is based
on an effective field theory where the (at present, in detail
unknown) quark and gluon degrees of freedom are parametrized by the bare
quantities of the effective Lagrangian. We assume here that QCD provides 
us with cutoff functions for the $N$-$N \pi$ vertices 
that make all integrations convergent. 
These vertices of the effective Lagrangian are ``bare'' at the hadronic level, 
i.e., they can still be dressed by purely hadronic mechanisms.

The physical currents for all processes contributing to the pion production 
amplitude are derived via their corresponding hadronic $n$-point Green's 
functions by employing a mathematical operation 
called a ``gauge derivative'' which allows one to obtain currents directly 
from the momentum-space versions of the respective Green's 
functions. For local fields, this is equivalent to the usual 
minimal-substitution procedure. However, for the present nonlocal case, 
where we assume bare 
vertex functions originating from QCD, this goes beyond minimal 
substitution. For lack of space, we cannot go into any details here 
regarding the gauge derivative. For the same reason, we rely entirely on 
a --- hopefully largely self-explanatory --- diagrammatic 
exposition of the formalism. Complete details 
are given in Ref. \cite{hh97}.

\section*{PION PHOTOPRODUCTION}

Generically, at any level of dynamical sophistication, there are four 
distinctly different contributions to the current $M^{\mu}$ for 
the pion photoproduction process $\gamma N\rightarrow \pi N$: Three 
contributions from the photon interacting with the 
three external hadronic legs of the $N$-$N \pi$ vertex and a fourth one
arising from the photon attaching itself within the vertex 
itself. At the simplest level where one has only (bare) tree graphs, 
this last contribution 
vanishes identically for a local pseudoscalar vertex; 
for a bare $N$-$N \pi$ vertex with derivative coupling, 
it is equal to the Kroll--Ruderman \cite{kroll54} contact term 
obtained from minimal 
substitution.

\begin{figure}[b!]
\centerline{\psfig{file=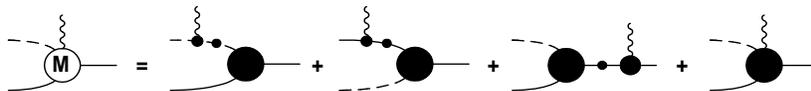,width=4.5in,clip=,silent=}}
\vspace{3mm}
\caption{\label{fig1} Basic diagrams describing pion 
photoproduction. Solid and dashed lines describe nucleons and pions, 
respectively.}
\end{figure}

\begin{figure}
\centerline{\psfig{file=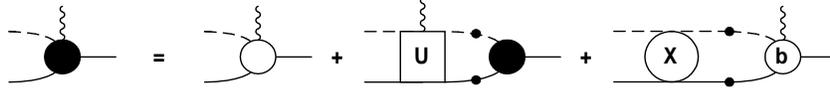,width=4.5in,clip=,silent=}}
\vspace{3mm}
\caption{\label{fig2} Interaction current. The first graph on the 
right-hand side is the bare contact term present already in the simplest
tree-level case.}
\end{figure}

\begin{figure}
\centerline{\psfig{file=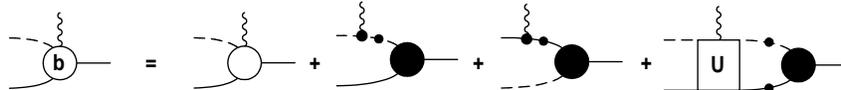,width=4.5in,clip=,silent=}}
\vspace{3mm}
\caption{\label{fig3} Auxiliary current $b^{\mu}$
 required in Fig.\ \ref{fig2}.
The exchange current $U^{\mu}$ appearing here and in Fig.\ \ref{fig2}
is shown in Fig.\ \ref{figU}.}
\end{figure}

\begin{figure}
\centerline{\psfig{file=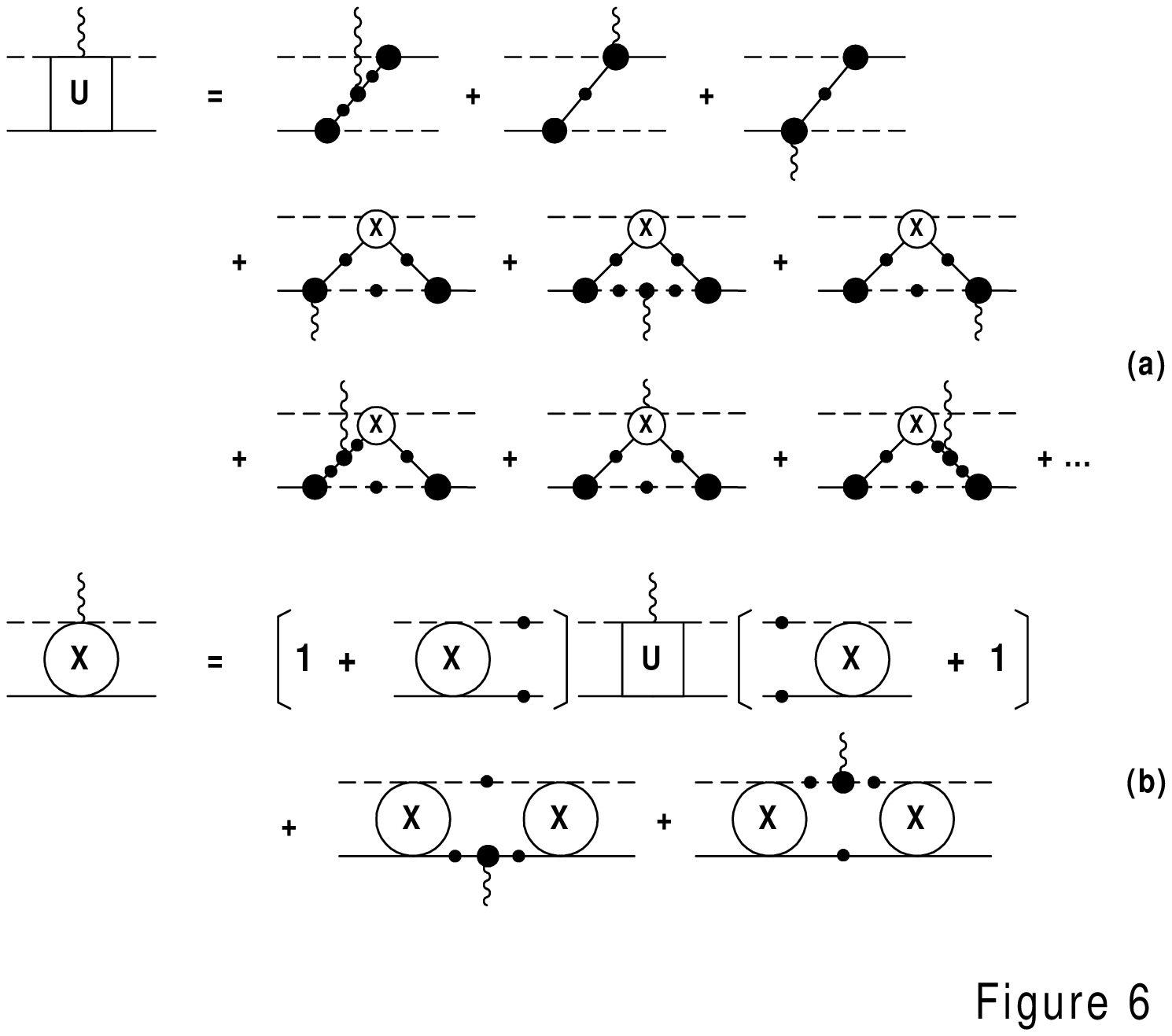,width=4.5in,clip=,silent=}}
\vspace{3mm}
\caption[title]{\label{figU} 
(a) Exchange current $U^{\mu}$ due to driving term $U$ and 
(b) current $X^{\mu}$ appearing in (a).}
\end{figure}

\begin{figure}
\centerline{\psfig{file=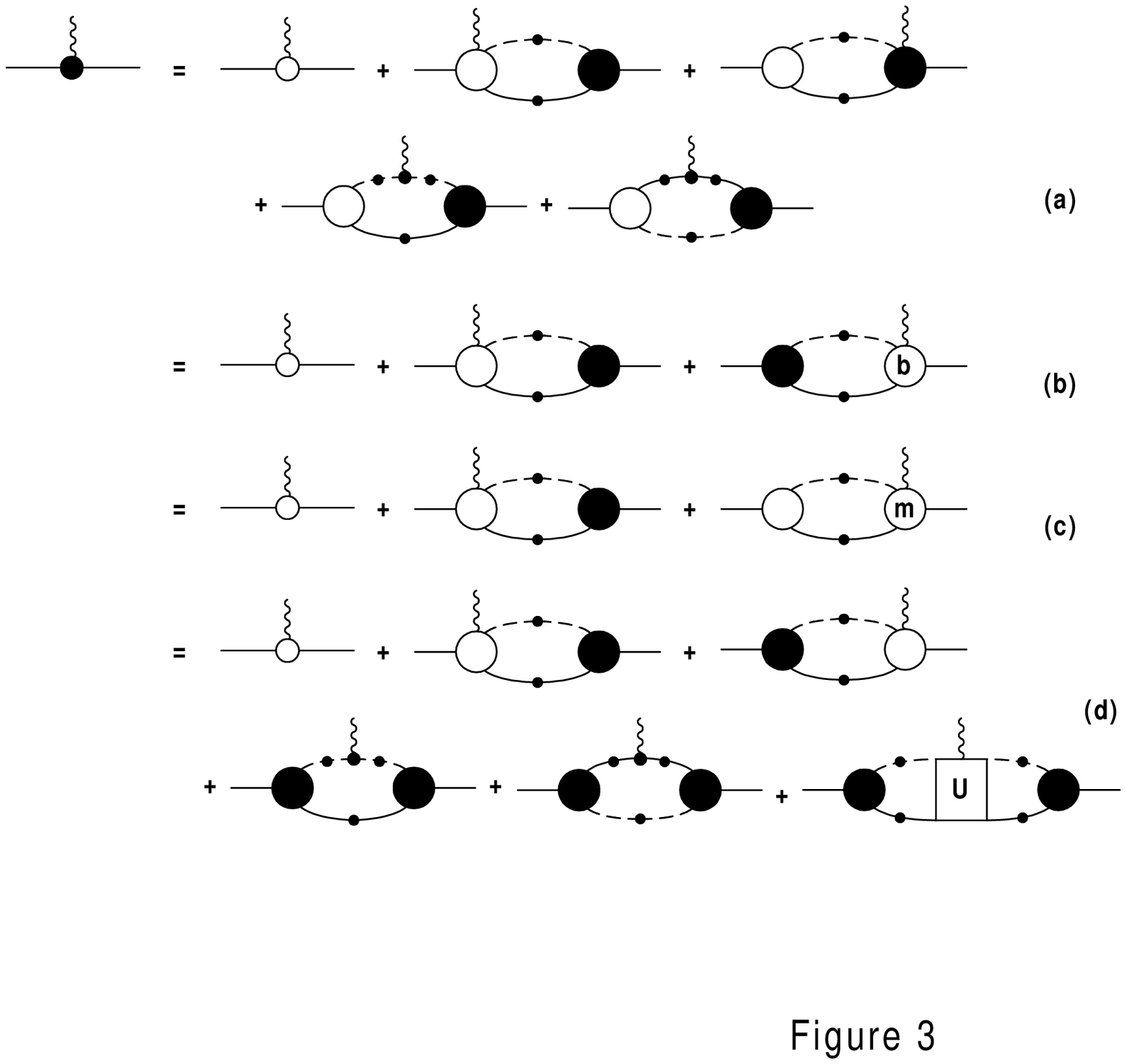,width=4.5in,clip=,silent=}}
\vspace{3mm}
\caption[title]{\label{figJ} 
Various equivalent representations of the fully dressed electromagnetic 
current for the nucleon.}
\end{figure}

This generic picture, which is shown in Fig.\ \ref{fig1}, 
remains true even if we consider fully dressed, physical hadrons. 
What becomes more complicated then are the ingredients contributing to the 
various mechanisms depicted in Fig.\ \ref{fig1}. For the fully dressed case,
the internal details of the right-most graph of Fig.\ \ref{fig1} ---  
now called the
{\it interaction current} (where, except for the bare contact term,
there is at least one hadronic $N$-$N \pi$ 
vertex before or after the photon interacts with the hadronic system) ---
are found \cite{hh97} to be given by the mechanisms shown in Fig.\ 
\ref{fig2}.
The interaction current is seen to contain a bare Kroll--Ruderman-type
contact term which is 
present already at the tree-level. In addition, it contains an exchange 
current contribution $U^{\mu}$ and an auxiliary current $b^{\mu}$ dressed 
by hadronic final-state interactions denoted by $X$. The details of 
$b^{\mu}$ are given in Fig.\ \ref{fig3}. 
The remaining current pieces are depicted in Figs.\ \ref{figU} and 
\ref{figJ}, 
where the latter shows the (on {\it and} off-shell) current for a
composite nucleon. 

\begin{figure}
\centerline{\psfig{file=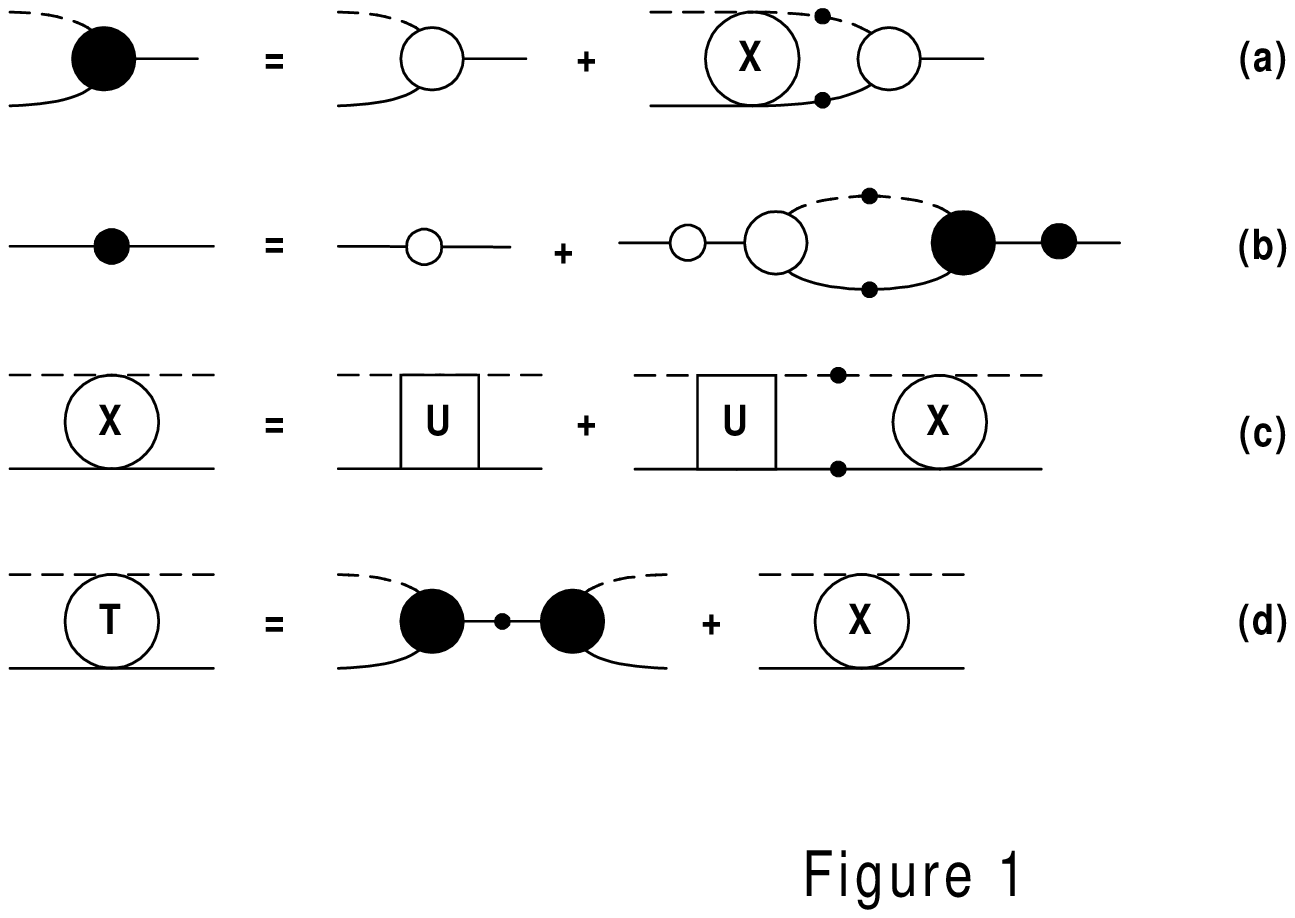,width=3in,clip=,silent=}\hfill%
\psfig{file=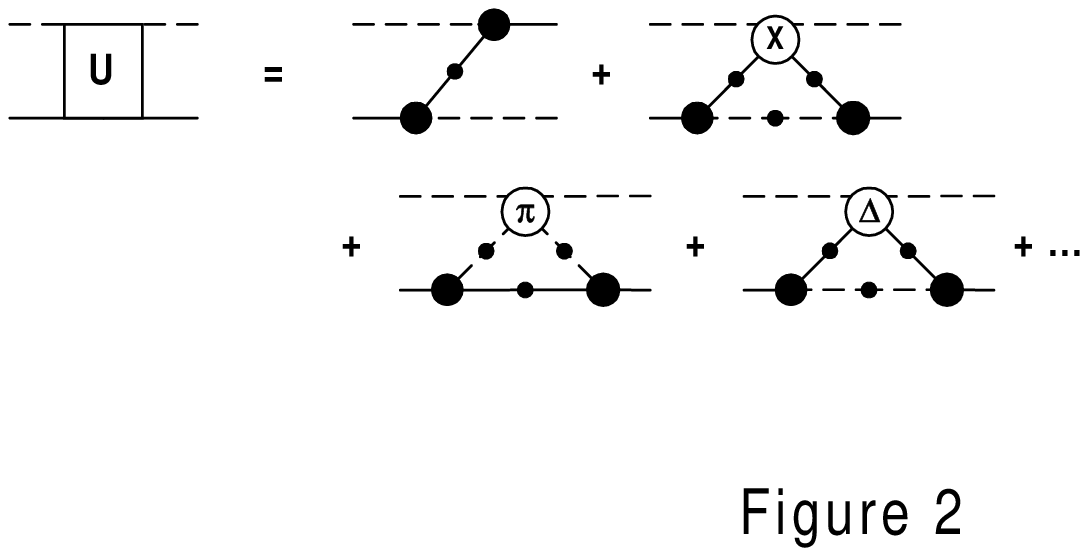,width=3in,clip=,silent=}} 
\vspace{3mm}
\caption[Summary of coupled, nonlinear equations for {$\pi N\rightarrow 
\pi N$}]%
{\label{fig4} %
Summary of coupled, nonlinear equations for {$\pi N\rightarrow 
\pi N$} in the $P_{11}$ channel \cite{hh97} (a similar set obtains
for the $P_{33}$ channel): 
(a) Fully dressed $N$-$N\pi$ vertex (filled circle) 
given in terms of bare vertex
(open circle) and final-state interaction mediated by the nonpolar $\pi N$
amplitude $X$. 
(b) Nonlinear equation for the dressed 
nucleon propagator (line with solid circle) determined by the bare
propagator (line with open circle) and the $\pi N$ self-energy bubble.
(c) Integral equation for the auxiliary amplitude $X$ given in terms of
the $\pi N$ irreducible driving term $U$ and intermediate propagation
of noninteracting $\pi N$ system.
(d) Full $\pi N$ Bethe--Salpeter amplitude $T$, expressed as the sum of the 
polar $s$-channel term (with
fully dressed $\pi N$ vertices and $N$ propagator) and the solution $X$
of (c).
Some of the lowest-order contributions for the driving term $U$ of
the nonpolar amplitude $X$ are given in the 
figure on the right-hand side. The corresponding first graph subsumes 
the $u$-channel exchanges of both $N$ and $\Delta$, 
i.e., it also incorporates
the crossing-symmetric partner of the $s$-channel 
graph of (d). The
other three box graphs depict intermediate scattering processes of the
nonpolar {$P_{11}$ $\pi N$} amplitude $X$ dressed by a pion, of the {$\pi
\pi $} amplitude dressed by a baryon, and of the full {$P_{33}$ $\pi N$}
amplitude with pion dressing, respectively.}
\end{figure}

As alluded to in the Introduction, the basis for this dynamical 
structure is the gauge-derivative formalism of Ref. \cite{hh97}. The 
underlying effective hadronic field theory describing $\pi N$ scattering, 
which supplies the required $n$-point Green's functions, is summarized in 
Fig.\ \ref{fig4}. 
The formalism is seen to be highly nonlinear at both the hadronic and 
electromagnetic levels.

\section*{GAUGE INVARIANCE}

\begin{figure}
\centerline{\psfig{file=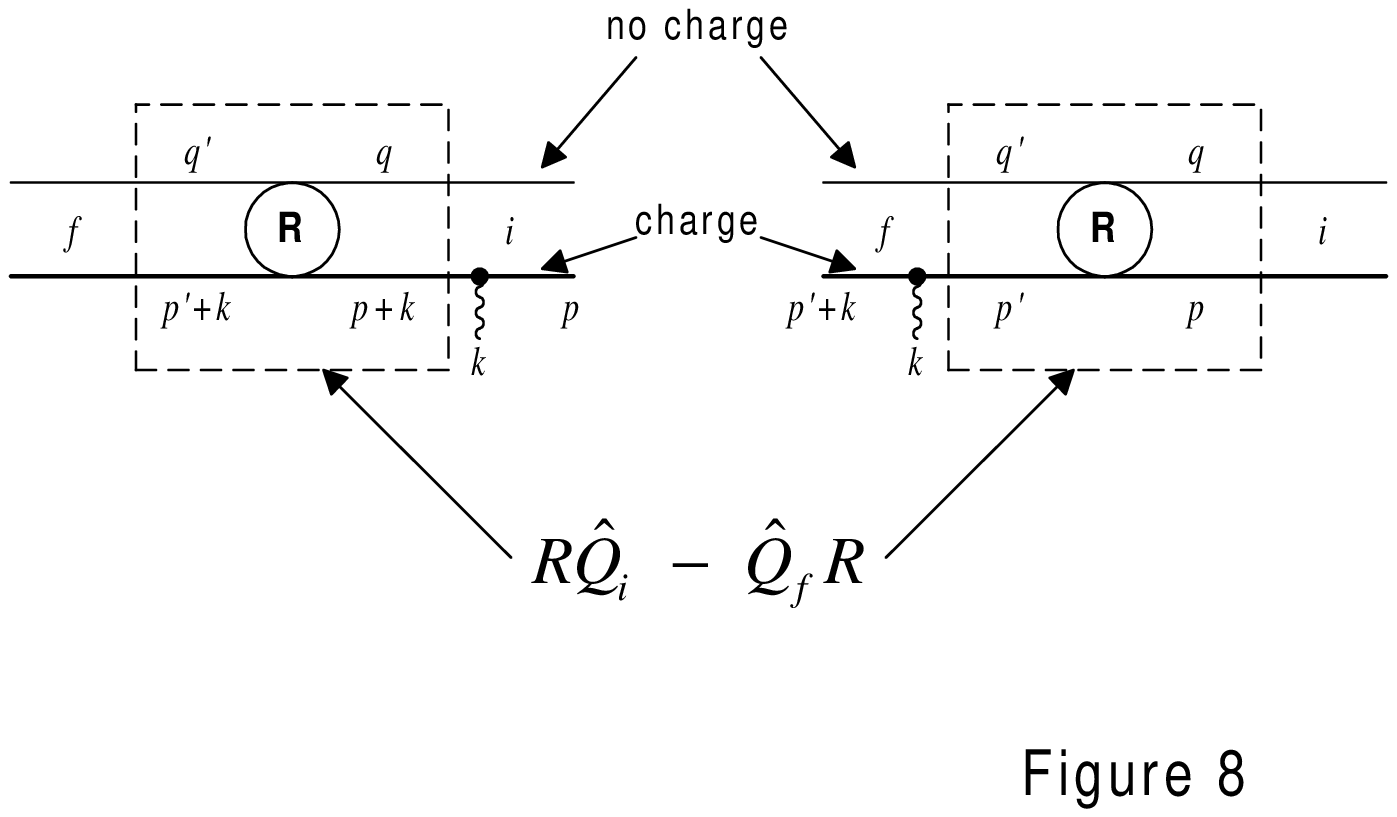,width=4.5in,clip=,silent=}}
\vspace{3mm}\caption[title]{\label{fgRmu} 
Generic representation of 
$R \widehat{Q}_i - \widehat{Q}_f R$  of Eq.\ (\ref{eq1}). 
$R$ is an arbitrary hadronic reaction mechanism where all incoming
and outgoing {\it uncharged} particles have been subsumed in the upper lines
and all incoming and outgoing {\it charged} particles in the lower, thicker
lines. The first graph on the left sums up all contributions where the
photon is attached to an incoming particle whereas the one on the right
depicts the sum of all contributions from the photon being attached to an
outgoing particle. 
$R \widehat{Q}_i - \widehat{Q}_f R$ is the difference between
the purely hadronic contributions enclosed in the dashed boxes; it measures
the change brought about in the hadronic reaction when a photon momentum is
transmitted through the hadronic interaction region entering and leaving via
charged particles.}
\end{figure}

The general condition for gauge invariance of all physcial currents
can be 
formulated very succinctly: All current contributions $R^{\mu}$
associated with the photon's interaction {\it within}
the interaction region of any hadronic reaction $R$ must satisfy the 
equation \cite{hh97}
\begin{eqnarray}
k_{\mu} R^{\mu} + R \widehat{Q}_i - \widehat{Q}_f R = 0 \; ,
\label{eq1}
\end{eqnarray}
where $Q_i$ is the
sum of all charge operators for all incoming
and $Q_f$ for all outgoing particle legs; the caret signifies that for all 
subsequent reaction mechanisms the particle described by the respective 
individual
charge operator will have the photon's four-momentum $k^{\mu}$ added to 
its own. The situation is illustrated in Fig.\ \ref{fgRmu}.
%
% === PSFIG ==================================== 
% ============================================== 
%
This is an {\it off-shell} generalization of the well-known  
Ward-Takahashi identity for 
propagators \cite{ward50,takah57} to the case of several incoming and outgoing 
particles. (Transverse magnetic moment contributions are contained in 
$R^{\mu}$; they are conserved by themselves, of course.)

As has been shown in Ref.\ \cite{hh97}, all reaction elements 
appearing in Figs.\ \ref{fig1}--\ref{figJ} 
do satisfy the gauge condition (\ref{eq1}) and therefore all physical
currents are indeed gauge-invariant. 

From a practical perspective, in view of the high nonlinearity of the 
present formalism, the question arises how this gauge invariance can be
maintained if some elements of the reaction mechanisms can only be
treated by approximations. 
Two well-known methods of ensuring gauge invariance in  
situations where the proper interaction currents 
are not available are due to Gross and 
Riska \cite{gross} and Ohta \cite{ohta}. The recent work of \cite{banerjee} 
suggests that there may be a theoretical
problem with Ohta's method having to do with
the fact that, in general, it requires knowledge of the 
vertex functions at unphysical values of its arguments which can be traced 
to an underlying non-Hermitian effective Lagrangian.

\begin{figure}
\centerline{\psfig{file=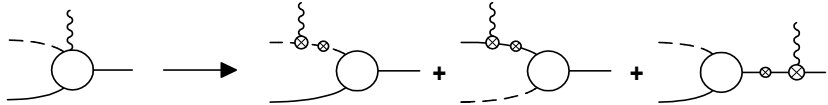,width=4.5in,clip=,silent=}}
\vspace{0mm}\caption[title]{\label{fgfmu} 
Effective treatment of the bare current as an example of the gauge-invariant
current approximation given in Eq.\ (\ref{eq2}), where the current
pieces due to a photon attaching itself within a hadronic interaction
region is replaced by a sum of suitably constructed currents over all
incoming and outgoing hadron legs. Each of the three leg currents is given
by an entire diagram on the right-hand side and can be interpreted 
\cite{hh97} as
a product of a particle current, a propagator, and a bare vertex 
function, in the manner shown here. For a constant vertex with derivative
coupling, the result is identical to the Kroll--Ruderman term 
\cite{kroll54} (see Eq. (93) of \cite{hh97}).}
\end{figure}

Haberzettl \cite{hh97} has suggested another, quite general,
method of maintaining gauge invariance when approximating interaction
currents which essentially replaces the full interaction current $R^{\mu}$ 
by a sum of single-particle currents over all incoming and outgoing hadron 
legs of the reaction $R$,
\begin{eqnarray}
R^{\mu} \rightarrow R^{\mu}_{\rm approx}
= \sum_{x_f} j^{\mu}_{{\rm leg},x_f} + \sum_{x_i} j^{\mu}_{{\rm leg},x_i} 
\;,
\label{eq2}
\end{eqnarray}
where the individual leg currents $j^{\mu}_{{\rm leg},x}$ are obtained
from a straightforward procedure using only charge conservation across 
the hadronic reaction in question (for details, see \cite{hh97}).
When applied to a nonlocal
vertex as, e.g., the bare $N$-$N \pi$ vertex (see Fig.\ \ref{fgfmu})
% === PSFIG ==================================== 
% ============================================== 
whose associated bare (QCD-based) interaction current is 
undetermined at the hadronic level, the resulting expressions are 
reminiscent of Ohta's \cite{ohta} for an extended
vertex. However, they are different in detail and
require only physical values of the vertex functions 
and, therefore, do not suffer from any of the problems discussed 
in Ref.\ \cite{banerjee}. We add that, of course, the approximate current 
(\ref{eq2}) can be amended by transverse pieces which are conserved by 
themselves.

\bibliographystyle{unsrt}

\end{document}